\documentclass[aps,prb,twocolumn,superscriptaddress,showpacs,amsmath,amssymb,longbibliography]{revtex4-2}
\usepackage{graphicx}
\usepackage[colorlinks=true,citecolor=blue,linkcolor=blue,urlcolor=blue,bookmarks=true]{hyperref}

\usepackage[T1]{fontenc}
\usepackage[utf8]{inputenc}
\usepackage[english]{babel}
\usepackage{amsmath}
\usepackage{xcolor}
\usepackage{soul}

\definecolor{nred} {RGB}{224,0,0}
\definecolor{nblue} {RGB}{28,130,185}
\definecolor{dgreen}{RGB}{78,138,21}
\definecolor{norange}{RGB}{230,120,20}

\begin{document}

\title{Local integrals of motion encoded in a few eigenstates}  
\author{J. Paw\l{o}wski}
\affiliation{Institute of Theoretical Physics, Faculty of Fundamental Problems of Technology, Wroc\l{a}w University of Science and Technology, 50-370 Wroc\l{a}w, Poland}
\author{P. Łydżba}
\affiliation{Institute of Theoretical Physics, Faculty of Fundamental Problems of Technology, Wroc\l{a}w University of Science and Technology, 50-370 Wroc\l{a}w, Poland}
\author{M. Mierzejewski}
\affiliation{Institute of Theoretical Physics, Faculty of Fundamental Problems of Technology, Wroc\l{a}w University of Science and Technology, 50-370 Wroc\l{a}w, Poland}

\begin{abstract}
Many properties of a quantum system can be obtained from just a single eigenstate of its Hamiltonian. For example, a single eigenstate can be used to determine whether a system is integrable or chaotic and, in the latter case, to establish its thermal properties. Focusing on the XXZ model, we show that the local integrals of motion, which lie at the heart of integrability, can also be estimated from a small number of eigenstates. Moreover, as the system size increases, fewer eigenstates are required, so that in the thermodynamic limit, the integrals of motion can be obtained from a vanishingly small fraction of all eigenstates. Interestingly, this property does not extend to integrals of motion arising solely from Hilbert space fragmentation, as found in the folded XXZ model, where the majority of eigenstates has to be used. This represents one of the few fundamental differences known between integrability and Hilbert space fragmentation.
\end{abstract}

\maketitle

\section{Introduction}  

Over the past two decades, significant effort has been devoted to understanding the non-equilibrium dynamics of closed quantum systems~\cite{rigol08,polkovnikov2011,dalessio16,mori18}. In particular, whether and, if yes, why the expectation values of local observables at long times agree with the predictions of standard Gibbs ensembles. This behavior is known as thermalization, while systems that exhibit it are commonly referred to as ergodic.

These studies have shown that a great deal of information can be extracted from just a single eigenstate in the thermodynamic limit, or from a small number of eigenstates in a finite system~\cite{garrison18}. For example, it can be used to identify ergodic behavior. The bipartite entanglement entropy of a single eigenstate, which quantifies correlations between subsystems, is universal in ergodic systems. Specifically, its leading term matches that of a random state from the Hilbert space~\cite{page93} and is maximal in the thermodynamic limit~\cite{beugeling15,vidmar17,haque22,bianchi22,kliczkowski23}. The only restriction is that the eigenstate cannot be too close to the edges of the spectrum~\cite{haque25}. For this reason, it has been used in studies of ergodicity-breaking transitions to pinpoint the transition point~\cite{bauer13,serbyn13,pawlik24,suntajs24,sierant25}.

Moreover, thermalization is currently understood through the Eigenstate Thermalization Hypothesis (ETH), which asserts that individual eigenstates are essentially thermal~\cite{srednicki99,deutsch18,steinigeweg14}. This means that the expectation values of local observables in a single eigenstate agree with the predictions of standard Gibbs ensembles. Consequently, a single eigenstate is not only sufficient to identify ergodicity, but also encodes the equilibrium properties of ergodic systems.

The most studied exceptions to thermalization are integrable models. Here, we focus on a paradigmatic example, the XXZ spin chain~\cite{yang66,yang66b,yang66c}. Integrable models support an extensive number of local integrals of motion (LIOMs), which are local operators that commute with the Hamiltonian, ensuring that their expectation values are conserved during time evolution. Integrable models do not satisfy the ETH, so their eigenstates are not thermal~\cite{haque14,alba15,leblond19,mierzejewski2020}. Nevertheless, in this work, we demonstrate that a small number of eigenstates suffices to determine LIOMs and, in principle, to construct the Generalized Gibbs Ensemble~\cite{rigol07,pozsgay13,ilievski15,vidmar16,kollar2011,cassidy2011,fukai2020}. Importantly, this number does not increase with system size, meaning that it represents a vanishing fraction of the Hilbert space dimension. This leads to the surprising conclusion that a small number of eigenstates can provide an accurate description of the stationary state also in an integrable model.
The identification of LIOMs also provides the starting point for constructing hydrodynamics or generalized hydrodynamics \cite{bertini16,castro-alvaredo16,Ilievski17,doyon2017,gopalakrishnan18,denardis18,agrawal20,friedman20,bastianello21,gopalakrishnan23,hubner25,doyon25}.

Finally, we verify whether this observation holds in systems exhibiting Hilbert space fragmentation~\cite{khemani17,sala2020,moudgalya21,moudgalya2022b,francica23,brighi23,sreemayee24,maitri2025}. In this ergodicity-breaking phenomenon, the Hamiltonian, when written in a suitable local basis, decomposes into exponentially many disconnected blocks. Each block contains an exponentially small fraction of all eigenstates. This phenomenon is also believed to be distinct from integrability. Although some fragmented systems are known to support LIOMs~\cite{lydzba2024}, it remains unclear whether all do. Furthermore, a seminal work connected Hilbert space fragmentation to a different class of conserved operators, which are not strictly local but behave as local in typical states from the Hilbert space~\cite{rakovszky20,moudgalya22}. Despite this, there are not many reported differences between integrability and Hilbert space fragmentation.

For this reason, we focus on the folded XXZ model, which can be obtained from the XXZ model in the limit of infinite anisotropy~\cite{pozsgay2021,zadnik21, zadnik21a}. We demonstrate that, apart from LIOMs stemming from Bethe-ansatz integrability (which vanish when a perturbation breaking Bethe-ansatz integrability is introduced), the model also hosts LIOMs originating from fragmentation (which do not vanish under such perturbations). We observe that the former can still be numerically generated from a small number of eigenstates, while the latter require almost all eigenstates. This leads to the conclusion that the information about LIOMs originating from fragmentation is encoded differently in eigenstates.

The paper is organized as follows. In Sec.~\ref{sec:finding}, we introduce the numerical method used to estimate LIOMs from an incomplete set of eigenstates. In Sec.~\ref{sec:comp}, we demonstrate that this estimation is accurate for a small number of eigenstates in the XXZ spin chain. We consider different symmetry sectors and extend our study to quasilocal integrals of motion. In Sec.~\ref{sec:frag}, we examine the effectiveness of the estimation in the folded XXZ model. Finally, we summarize our findings and conclude in Sec.~\ref{sec:con}.

\section{Finding LIOMs from incomplete spectrum}
\label{sec:finding}

We first discuss the numerical method for estimating LIOMs, denoted hereafter as $\hat {\cal Q}^{\alpha}$, from an incomplete set of eigenstates of the Hamiltonian, \mbox{\(  H |n\rangle=E_n|n\rangle \)}. Here, the index $\alpha$ labels LIOMs. We focus on Hermitian operators, for which we use the Hilbert-Schmidt (HS) product 
$\langle A   B \rangle ={\rm Tr} ( A   B )/Z$ and the HS norm $\lVert A\lVert^2=\langle A  A \rangle $, where the trace is calculated over the {\em entire} Hilbert space with dimension $Z$. We require that $||A||^2=1$.
For simplicity, we first consider a nondegenerate spectrum. This assumption is not essential, and minor modifications allow to study systems with degenerate eigenvalues. This is discussed in sections~\ref{sec:comp} and~\ref{sec:frag}.
Note that, in a case of nondegenerate  spectrum, conserved quantities have nonvanishing matrix elements only along the diagonal, i.e., $\langle m|{\cal Q}^{\alpha}|n \rangle \ne 0$ only for $m=n$. Therefore, in order to confirm that $\mathcal{Q}_{\alpha}$ is conserved, one needs to calculate at least all diagonal matrix elements and verify that $1/Z \sum_n \langle n|{\cal Q}^{\alpha}|n \rangle^2=\lVert {\cal Q}^{\alpha} \lVert^2 $. This task requires full diagonalization of the Hamiltonian, posing a bottleneck for numerical studies.

In order to estimate LIOMs from selected eigenstates, it is necessary to implement a method that does not rely on the HS product or HS norm (unlike the one employed in~\cite{mierzejewski15b,mierzejewski15c,mierzejewski18}). An appropriate approach has recently been introduced in Ref.~\cite{lydzba2024}. It is based on the observation that LIOMs can be obtained from compressing the information stored in the diagonal matrix elements of local operators, $A^s_{nn}=\langle n |\hat A^s | n\rangle$. By local operators, we mean (sums of) operators that act nontrivially on a finite subvolume of a system in the thermodynamic limit. To this end, we consider a set of $D_O$ orthonormal local operators
$\{A^1,\ldots,A^{D_O}\}$, which satisfy $\langle A^i A^j \rangle = \delta_{ij}$. Next, we construct a matrix of dimension $Z\times D_O$ composed of their diagonal matrix elements,
\begin{equation}
    {\cal R}=
\begin{bmatrix}
A_{11}^{1} & A_{11}^{2} & . & . & . & A_{11}^{D_O}\\ 
A_{22}^{1} & A_{22}^{2} & . & . & . & A_{22}^{D_O}\\ 
. & . & . & . & . & .\\
. & . & . & . & . & .\\
. & . & . & . & . & .\\
A_{ZZ}^{1} & A_{ZZ}^{2} & . & . & . & A_{ZZ}^{D_O} 
\end{bmatrix}\;,
\label{rcal}
\end{equation}
and perform its singular value decomposition, 
\mbox{${\cal R}=\mathcal{U} \tilde{\Lambda}\mathcal{V}^{T} $}. Here, $\mathcal{U}$ and $\mathcal{V}$ are orthogonal matrices of dimensions $Z\times Z$ and $D_O\times D_O$, respectively. Additionally, \mbox{$\tilde{\Lambda}=\mathrm{diag}[\tilde{\lambda}_1,\tilde{\lambda}_2,\ldots ,\tilde{\lambda}_{{\rm min}(D_O,Z)}]$}, and its elements on the main diagonal are the so-called singular values, while all other elements vanish. The rank of $\mathcal{R}$ is at most ${\rm min}(D_O,Z)$, and in practice $D_O < Z$. Numerically, we always perform the so-called thin SVD instead of the full SVD, in which only the first $D_O$ columns of the matrix $\mathcal{U}$ are retained, while $\tilde{\Lambda}$ becomes a square matrix of dimension $D_O\times D_O$.
The compression of ${\cal R}$ consists of finding a matrix $\tilde{\cal R}$
with a smaller rank that is as close as possible to ${\cal R}$ (formally, that minimizes $\lVert {\cal R}-\tilde{\cal R} \rVert$). According to the Eckart–Young–Mirsky theorem, this compression amounts to preserving the largest singular values and discarding the others. The number of preserved singular values sets the rank of $\tilde{\cal R}$. It can also be demonstrated~\cite{lydzba2024}, that the largest possible singular values, $\tilde{\lambda}_{\alpha}=\sqrt{Z}$, correspond to LIOMs, ${\cal  Q}^{\alpha}$. These can be expressed in terms of the local operators introduced in Eq.~\eqref{rcal}, i.e., ${\cal Q}^{\alpha}=\sum_{s=1}^{D_O} \mathcal{V}_{s\alpha}A^{s}$. In other words, the compression carried out for the initial set of operators $\{A^1,\ldots, A^{D_O} \}$ finds all LIOMs that can be written as linear combinations of these operators. 

In this work, we show that the LIOMs can be accurately estimated by compressing a small number of randomly selected rows of ${\cal R}$. 
While restricting the compression to a subset of eigenstates does not guarantee finding the true LIOMs, the problem remains well-posed, as it does not rely on the HS product.
Specifically, we randomly select $N_{S}$ eigenstates of the Hamiltonian that correspond to $N_{S}$ rows of the matrix ${\cal R}$. These rows are then stored in the matrix $R$, for which we perform the singular value decomposition:
\begin{equation}
R=\begin{bmatrix}
A_{11}^{1}  & . & . & . & A_{11}^{D_O}\\ 
. & . & . & . & .\\
. & . & . & . & .\\
. & . & .  & . & .\\
A_{N_S N_S}^{1} & . & . & . & A_{N_S N_S}^{D_O} 
\end{bmatrix}=U {\rm diag}[\lambda_1,\lambda_2,\ldots] V^T \;.
\label{svd}
\end{equation}
Next, we define the approximate LIOMs as
\begin{equation}
Q^{\alpha}=\sum_{s=1}^{D_O} V_{s\alpha}A^{s} \;,  
\label{liom}
\end{equation}
for the largest singular values $\lambda_{\alpha}$. Moreover, we demonstrate that these approximate LIOMs are very close to the true LIOMs, $Q^{\alpha}\simeq {\cal Q}^{\alpha}$, for an unexpectedly small number of eigenstates, $N_{S}\ll Z$. We emphasize that, from now on, calligraphic letters denote quantities obtained from the full set of eigenstates, e.g., $\mathcal{R}$, while non-calligraphic letters denote corresponding quantities obtained from a subset of eigenstates, e.g., $R$.

The selected eigenstates are numerically obtained using the shift-invert method for the target energy $E_{T}$~\cite{pietracaprina18}. Since we focus on the translationally-invariant spin chains that conserve the total spin projection, we randomly choose $N_S$ values of $E_T$, distributed across all considered magnetization and momentum sectors according to the appropriate binomial and uniform distributions, respectively.
In each symmetry sector, the target energies are sampled with the probability density equal to the many-body density of states, $\rho(E_T)= \frac{1}{Z}\sum_{n}\delta(E_T-E_n)= \exp[-E^2_T/(2 L\sigma^2)]/\sqrt{2\pi L \sigma^2 }$, which variance is obtained from the exact diagonalization of small systems with $L\in\{14,16,18,20\}$. In this way, each eigenstate $|n \rangle$ has equal probability of being selected.

\section{Compression-based approach in XXZ model}
\label{sec:comp}

As mentioned above, we consider the translationally-invariant spin models that preserve the total spin projection, \mbox{$S^z_{tot}=\sum_l S^z_l$}, where $S_l$ are spin $s=\frac{1}{2}$ operators acting on site $l$.
For this reason, we study operators that are also translationally-invariant, Hermitian, and commute with $S^z_{tot}$. We group these operators according to their support, $M$, i.e., the maximal number of consecutive sites on which they act nontrivially. For example, 
$\sum_l S^z_l S^z_{l+1}$ and $\sum_l S^+_l S^z_{l+1} S^z_{l+2} S^-_{l+3} +{\rm H.c.} $ both belong to a group with $M=4$,
whereas the former one also belongs to a group with $M=2$.

We first establish the number of eigenstates required for finding LIOMs in the XXZ chain:  
\begin{equation}
H=\frac{1}{2} \sum_l \left( S^+_l S^-_{l+1}+S^-_l S^+_{l+1}\right)+
\Delta  \sum_l  S^z_l S^z_{l+1},
\label{xxz}
\end{equation} 
with $\Delta=3/4$ used in all calculations. We consider two orthogonal sectors of operators constructed as
\begin{eqnarray}
A_{R}=O+O^{\dagger}, \label{ro} \\
A_{I}=i(O-O^{\dagger}) \label{io},
\end{eqnarray}
which for brevity will be referred to as {\em real} and {\em imaginary} operators, respectively. 

Here, $O$ are Pauli strings constructed from products of operators acting on individual sites. At each site $l$, we consider the set $\{S_l^+, S_l^-, S_l^z, I_l\}$, where $I_l$ is identity operator acting on site $l$. From all possible operators $O$, we retain only these that commute with $S^z_{tot}$. The resulting operators $A_R$ and $A_I$ are made traceless and orthonormal.

It is worth noting that the XXZ chain hosts degeneracies. Two degenerate eigenstates either belong to different momentum sectors or both belong to the $k=0$ or $k=\pi$ momentum sectors. Since the considered operators, $A_R$ and $A_I$, are translationally invariant, they have no matrix elements between eigenstates from different momentum sectors. For this reason, we exclude the $k=0$ and $k=\pi$ momentum sectors from our numerical simulations.

Nevertheless, it is possible to handle degeneracies (if not massive). To this end, the shift-invert procedure for each target energy, $E_T$, has to be run multiple times. Since we start from a new random state in each run, if the closest eigenvalue to $E_T$ is degenerate, the procedure will converge to random non-orthogonal states from the degenerate subspace. On these states, we perform the SVD, and obtain the dimension of the degenerate subspace and its orthogonal basis. This allows to correctly compute all matrix elements in this degenerate subspace, and include them in the $R$ matrix, as explained in Ref.~\cite{lydzba2024}.

Using the algorithm described above, we have verified that consistent results are obtained if the $k=0$ and $k=\pi$ momentum sectors are taken into account (not shown). Moreover, the problem of massive degeneracies is addressed in the section devoted to the fragmented systems, i.e., Sec.~\ref{sec:frag}.

\subsection{Local integrals of motion}

We start with a compression problem for imaginary operators supported on up to $M=4$ sites that are even under the spin-flip transformation $S^{\pm}_l \to S^{\mp}_l$ and $S^z_l\to -S^z_l$. This sector contains $D_O=9$ operators, which permits the construction of only a single LIOM that corresponds to the energy current~\cite{zotos97,grabowski95}, 
\begin{equation}
{\cal Q}^1=\sum_{s=1}^3 {\cal V}_{s1} A^s, \label{encu}
\end{equation}
where 
\begin{eqnarray}
A^1 &=& i \sum_l S^+_l S^z_{l+1} S^-_{l+2} +{\rm H.c.}, \label{s11}\\
A^2 &=& i \sum_l S^z_l S^-_{l+1} S^+_{l+2} +{\rm H.c.}, \label{s112} \\
A^3 &=& i \sum_l S^-_l S^+_{l+1} S^z_{l+2} +{\rm H.c.}, \label{s12}
\end{eqnarray}
while ${\cal V}_{11}\simeq 0.6860$ and ${\cal V}_{21}={\cal V}_{31}\simeq 0.5145$, related by \(\mathcal{V}_{21}/\mathcal{V}_{11} = \mathcal{V}_{31}/\mathcal{V}_{11} = \Delta\).

\begin{figure}[t]
    \centering
    \includegraphics[width=\linewidth]{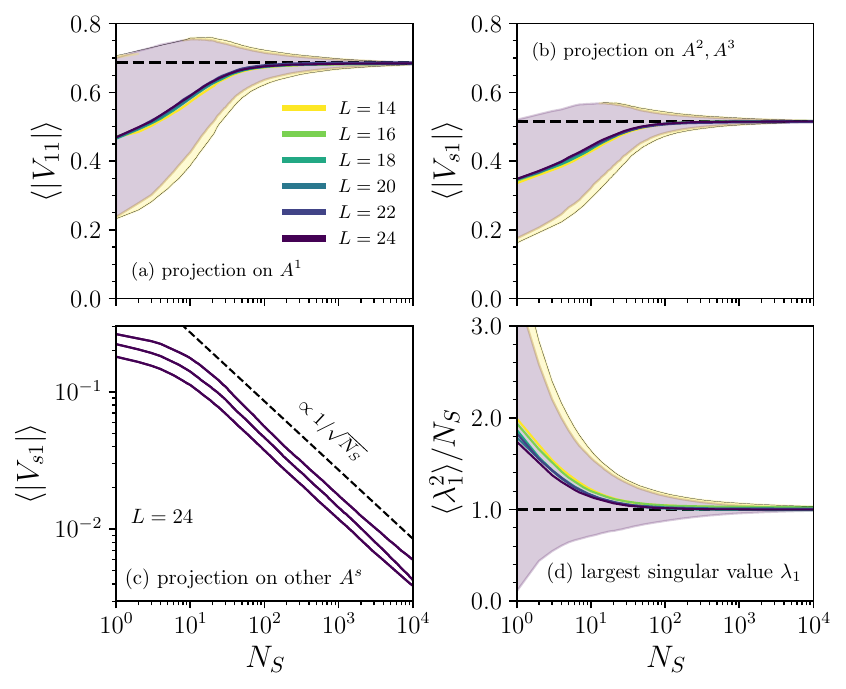}
    \caption{Numerical results for the largest singular value obtained for the set of imaginary operators that are even under the spin-flip transformation and supported on up to $M=4$ sites. This set $\{A^1,\ldots, A^{D_O} \}$
    contains $D_O=9$ operators including those in Eqs.~\eqref{s11}-\eqref{s12}.
    Continuous lines in (a) show the  projection $\langle |V_{11}| \rangle$ averaged over different choices of $N_s$ eigenstates for
    \(L=14,\ldots, 24\). The number of these choices is $10^4$, also in Figs.~\ref{fig2}-\ref{fig4}. The shaded areas indicate the standard deviation of $|V_{11}|$ for the smallest ($L=14$) and largest ($L=24$) systems, using consistent color coding. The edge of the former is marked with a thin black curve for visibility.
    Additionally, the horizontal dashed line marks the analytical prediction, ${\cal V}_{11}$, for the energy current, ${\cal Q}_1$, see Eq.~(\ref{encu}). (b)~The same resuts as in (a), but for the projection $V_{21}$. The projection $V_{31}$ is indistinguishable from $V_{21}$ on the presented scale. (c) Continuous lines show the projections $\langle |V_{s1}| \rangle$ on other operators $(s\ge 4)$ for \(L=24\).  
    (d) The largest singular value, $\langle \lambda^2_1\rangle/N_S$, see Eq.~(\ref{svd}). Strict LIOMs correspond to $\lambda^2/Z=1$ obtained from all eigenstates (when $N_S=Z$).  
    }
    \label{fig1}
\end{figure}

Figures~\ref{fig1}(a)-\ref{fig1}(c) show the average projection
$\langle | V_{s1} | \rangle$, as defined in Eq.~(\ref{liom}), which corresponds to the largest singular value $\lambda_1$ of the matrix $R$, see Eq.~(\ref{svd}). The averaging is performed over different choices of $N_S$ eigenstates of the Hamiltonian. We note that the overall sign of LIOMs ($Q^{\alpha}$ versus $-Q^{\alpha}$) is arbitrary and emerges randomly in the numerical procedure. For this reason, the averaging is carried out over the absolute values. The continuous lines in panels (a) and (b) show the average values $\langle | V_{11} | \rangle$ and  $\langle | V_{21} | \rangle=\langle | V_{31} | \rangle$, respectively. The shaded areas indicate the corresponding standard deviations obtained for the smallest ($L=14$) and largest ($L=24$) systems, using consistent color coding. The edge of the former is marked with a thin black curve for visibility.

It is apparent that the the energy current (LIOM from Eq.~\eqref{encu}) can be accurately estimated already from $N_S=N^*\simeq 10 \div 10^2$ eigenstates. The average projections, $\langle | V_{11} | \rangle$ and $\langle | V_{21} | \rangle=\langle | V_{31} | \rangle$, become very close to the exact results, ${\cal V}_{11}$ and ${\cal V}_{21}={\cal V}_{31}$. The latter are marked with dashed lines in Figs.~\ref{fig1}(a)-\ref{fig1}(b). At the same time, the average projections on other operators, $\langle | V_{s1} | \rangle$ with $s\ge 4$, decrease as $1/\sqrt{N_S}$, see Fig.~\ref{fig1}(c). This is accompanied by the rescaled singular value, $\langle \lambda^2_1\rangle/N_S$, saturating to one for $N_S=N^*$, see Fig.~\ref{fig1}(d). Not only do the average projections converge to the exact results, but their standard deviations also become very small. This implies that a randomly selected set of $N_S > N^*$ eigenstates allows to reconstruct the LIOM. Interestingly, the number of eigenstates needed for this purpose is almost independent of the system size or even slowly decreases with increasing $L$. The same finite-size scaling is observed for the standard deviations.
Simultaneously, the dimension of the Hilbert space is exponential in $L$.
Therefore, we can conclude that LIOMs can be obtained from an exponentially small fraction of eigenstates (e.g., $N_S=100$ for $L=24$ corresponds to only $\sim 10^{-5}$ of all eigenstates).

In order to explain what determines the minimal number of eigenstates, $N^*$, we have repeated the same calculations as in Fig.~\ref{fig1} but for a much larger set of local operators. To this end, we have expanded the support from $M=4$ to $M=6$, so that the compression problem is performed for the set of $D_O=155$ operators. This set allows to reconstruct two LIOMs: the previously discussed ${\cal Q}^1$ and an additional ${\cal Q}^2$ corresponding to the second largest singular value $\lambda_2$. The latter can be written as ${\cal Q}^2=\sum_{s=1}^{21} {\cal V}_{s2} A^s$, where the two projections with largest magnitude are on
\begin{eqnarray}
A^1 &=& i \sum_l S^+_{l} S^z_{l+1} S^z_{l+2} S^z_{l+3} S^-_{l+4} +{\rm H.c.}\;, \label{s21} \\
A^2 &=& i \sum_l S^z_{l} S^+_{l+1} S^z_{l+2} S^z_{l+3} S^-_{l+4} + {\rm H.c.}\;, \\
A^3 &=& i \sum_l S^+_{l} S^-_{l+1} S^+_{l+2} S^z_{l+3} S^-_{l+4} + {\rm H.c.}\;, \\
A^4 &=& i \sum_l S^-_{l} S^+_{l+1} S^+_{l+2} S^z_{l+3} S^-_{l+4} +{\rm H.c.}\;, \\
A^5 &=& i \sum_l S^+_{l} S^z_{l+1} S^-_{l+2} S^+_{l+3} S^-_{l+4} + {\rm H.c.}\;, \\
A^6 &=& i \sum_l S^-_{l} S^z_{l+1} S^+_{l+2} S^+_{l+3} S^-_{l+4} +{\rm H.c.}\;, \\
A^7 &=& i \sum_l S^+_{l} S^z_{l+1} S^z_{l+2} S^-_{l+3} S^z_{l+4} + {\rm H.c.}\;, \\
A^{8} &=& i \sum_l S^+_{l} S^z_{l+2} S^-_{l+3} + {\rm H.c.}\;, \\
A^{9} &=& i \sum_l S^z_{l} S^+_{l+2} S^-_{l+3} + {\rm H.c.}\;, \\
A^{10} &=& i \sum_l S^+_{l} S^z_{l+1} S^-_{l+3} + {\rm H.c.}\;, \\
A^{11} &=& i \sum_l S^+_{l} S^-_{l+1} S^2_{l+3} + {\rm H.c.}\;, \label{s22}
\end{eqnarray}
with values $\mathcal{V}_{12}\simeq 0.3344$, $\mathcal{V}_{22} = \mathcal{V}_{42} = \mathcal{V}_{72} = \mathcal{V}_{82} = \mathcal{V}_{10,2}\simeq -0.2508$ and 
 $\mathcal{V}_{32} = \mathcal{V}_{52} = \mathcal{V}_{62} = \mathcal{V}_{92} = \mathcal{V}_{11,2}\simeq 0.2508$.

\begin{figure}[t]
    \centering
    \includegraphics[width=\linewidth]{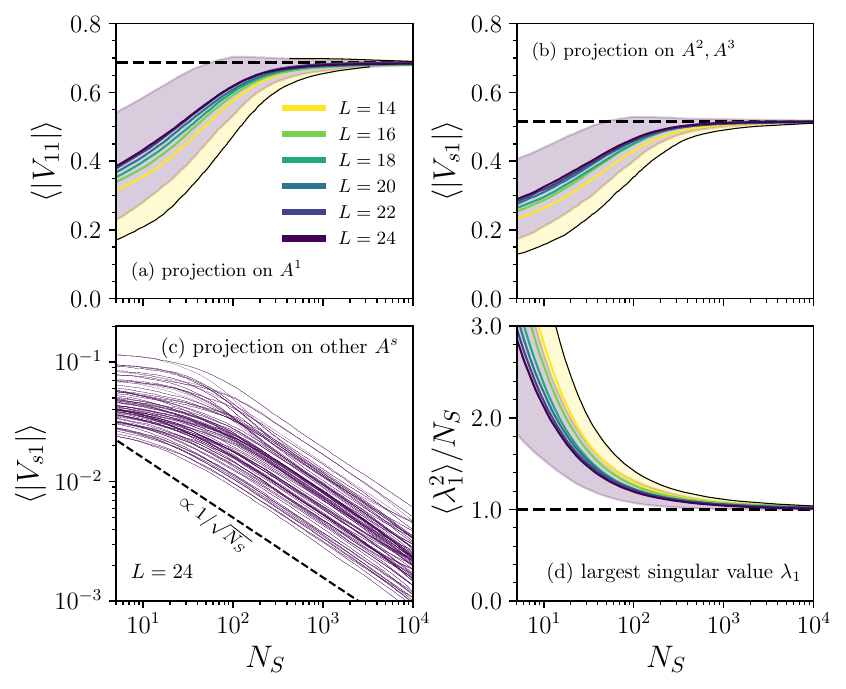}
    \caption{The same as in Fig.~\ref{fig1}, but for a larger support $M=6$, for which the set $\{A^1,\ldots, A^{D_O}\}$ contains $D_O=155$ operators. The approximate LIOMs corresponding to two largest singular values, $Q^{'1}$ and  $Q^{'2}$, are rotated according to the procedure described in the main text, see Eq.~\eqref{rot}.
    (a) and (b) show the projections of $Q^1$ on the operators from Eqs.~\eqref{s11} and~\eqref{s112}, respectively. The projections on operators from Eqs.~\eqref{s112}-\eqref{s12} are mutually indistinguishable on the presented scale. (c) The projections of $Q^1$ on other operators with $s\ge 4$. Dashed lines in (a) and (b) mark the analytical predictions, as explained below Eq.~\eqref{s12}.}
    \label{fig2}
\end{figure}
\begin{figure}[thbp]
    \centering
    \includegraphics[width=\linewidth]{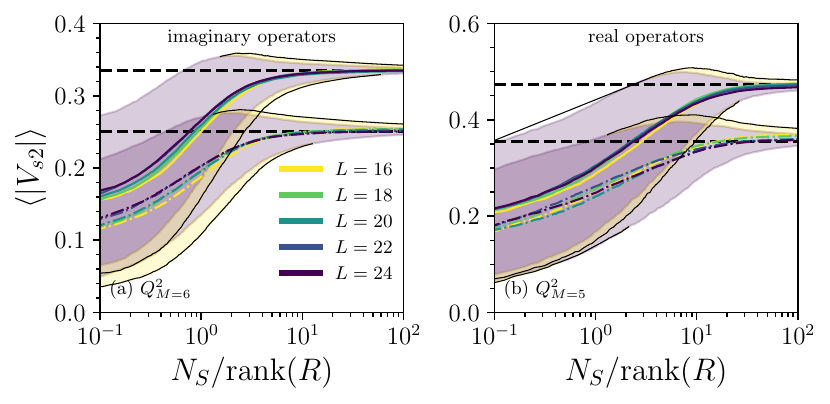}
    \caption{(a) The same as in Figs.~\ref{fig2}(a) and \ref{fig2}(b) but for \(Q^2\) instead of \(Q^1\). 
    This panel shows the projections of \(Q^2\) on the operators from Eqs.~\eqref{s21}-\eqref{s22}. 
    The largest projection is on $A^1$ from Eq.\eqref{s21}, while the projections on other operators are mutually indistinguishable on the scale shown in (a).
    The horizontal dashed lines mark the analytical predictions, see text below Eq.~\eqref{s22}. The number of eigenstates $N^*$ is rescaled by
    \( \mathrm{rank}({\cal R}) = 57\). (b) The numerical results for $Q^2$ constructed from the set of real operators that are even under the spin-flip transformation and are supported on up to $M=5$ sites. In this sector, $Q^1$ represents the normalized Hamiltonian. This plot shows the projections of $Q^2$ on the operators from Eqs.~\eqref{s31}-\eqref{s32}. The largest projection is on $A^1$ from Eq.~\eqref{s31} and the projections on other operators are indistinguishable on the scale shown in (b). The number of eigenstates $N^*$ is rescaled by \(\mathrm{rank}(R) = 31\).
    }
    \label{fig3}
\end{figure}

The presence of more than one LIOM introduces a minor technical complication, making it difficult to compare the numerical results with the known analytical formulas. Note that an arbitrary rotation of two orthonormal LIOMs for $M=6$, ${\cal Q}^1_{M}$ and ${\cal Q}^2_{M}$, generates another pair of orthonormal LIOMs, ${\cal Q'}^1_{M}$ and ${\cal Q'}^2_{M}$. They are related by
${\cal Q}^1_{M}=\cos(\phi)\,{\cal Q'}^1_M+\sin(\phi)\,{\cal Q'}^2_M$ and ${\cal Q}^2_M=\cos(\phi)\,{\cal Q'}^2_M-\sin(\phi)\,{\cal Q'}^1_M$.
The singular value decomposition gives an arbitrarily rotated pair of LIOMs, ${\cal Q'}^1_{M}$ and ${\cal Q'}^2_{M}$. The latter operators are also valid LIOMs, as they commute with the Hamiltonian and give the same contribution to the Mazur bound for any observable $O$, i.e., $\sum_{i=1}^{2}\langle O \mathcal{Q}^{'i}_{M}\rangle^2=\sum_{i=1}^{2}\langle O \mathcal{Q}^{i}_{M}\rangle^2$, which provides a lower bound for the autocorrelation function of $O$~\cite{mazur69,zotos97,dhar21}.
Nevertheless, in known analytical formulas it is always assumed that ${\cal Q}^1_{M} = {\cal Q}^1_{M-2}$, and so it is convenient to rotate the LIOMs obtained for a larger support such that they match the LIOMs obtained for a smaller $M$.
 Formally, this amounts to maximizing $\langle {\cal Q}^1_{M} {\cal Q}^1_{M-2} \rangle^2$, which is achieved for the rotation angle satisfying
\begin{equation}
\tan(2 \phi)=\frac{2 \langle {\cal Q'}^1_{M} {\cal Q}^1_{M-2}  \rangle  \langle {\cal Q'}^2_{M} {\cal Q}^1_{M-2} \rangle }{\langle {\cal Q'}^1_{M} {\cal Q}^1_{M-2}  \rangle^2-\langle {\cal Q'}^2_{M} {\cal Q}^1_{M-2} \rangle^2}. \label{rot}
\end{equation}

The same rotation can be performed for the approximate LIOMs 
from Eq.~\eqref{liom}. Namely, for each set of $N_S$ eigenstates, we have determined \( Q^1_{M-2}\) as well as a pair \( Q'^1_{M}\) and  \( Q'^2_{M}\), which is then rotated according to the procedure described above. As a result, we have obtained two approximate LIOMs: \( Q^1_{M}\simeq Q^1_{M-2} \)
and an additional \( Q^2_{M}\).  Figures~\ref{fig2} and~\ref{fig3}(a) show results for $Q^1=Q^1_{M}$  and  $Q^2=Q^2_{M}$, respectively.

We observe that the conclusions drawn from Fig.~\ref{fig1} remain valid for larger supports ($M=6$) and more complicated LIOMs ($\mathcal{Q}^2$). However, the more operators are considered, the more eigenstates are needed. In other words, $N^*$ increases with $M$. By comparing the numerical results for several $M$, we find that the estimation of LIOMs is effective when the ratio $N^*/{\rm rank}({\cal R})$ is of the order of ${\cal O}(1)\div{\cal O}(10)$. In Fig.~\ref{fig1}, $M=4$ and ${\rm rank}({\cal R})=5$, while in Fig.~\ref{fig2}, $M=6$ and ${\rm rank}({\cal R})=57$. The relation between $N^*$ and ${\rm rank}({\cal R})$ is consistent with the fact that the LIOMs are obtained from a compression that reduces the rank of $R$. For $N_{S}<{\rm rank}({\cal R})$, the rank is limited by the number of eigenstates, $N_S$. The compression then determines which states are the most relevant for the information contained in the matrix $R$ and need to be preserved while reducing its rank. Only for $N_{S}>{\rm rank}({\cal R})$ is the rank limited by the number of operators, $D_O$. In this case, the compression selects the most relevant operators, i.e., the LIOMs.
Therefore, the fact that $N^* >{\rm rank}(R)$ is not very surprising. What is, however, surprising is that the estimation of LIOMs becomes effective very soon after exceeding the threshold $N^*={\rm rank}({\cal R})$, i.e., soon after the compression starts to targets operators instead of states.

Finally, we confirm that these findings hold in other symmetry sectors. Figure \ref{fig3}(b) shows results of an analogous study, but for the real operators that are even under the spin-flip transformation and are supported on up to $M=5$ sites. This allows to reconstruct two LIOMs. Specifically, the Hamiltonian ${\cal Q}^1=H/\lVert H \rVert$
and an additional LIOM ${\cal Q}^2$, which can be written as ${\cal Q}^2=\sum_{s=1}^{10} {\cal V}_{s2} A^s$, where the projections with two largest magnitudes are on
\begin{eqnarray}
A^1 &=& \sum_l S^+_{l} S^z_{l+1} S^z_{l+2} S^-_{l+3} +{\rm H.c.}\;, \label{s31} \\
A^2 &=& \sum_l S^+_{l} S^-_{l+2} +{\rm H.c.}\;, \\
A^3 &=& \sum_l S^+_{l} S^z_{l+1} S^-_{l+2} S^z_{l+3}+{\rm H.c.}\;, \\
A^4 &=& \sum_l S^-_{l} S^+_{l+1} S^+_{l+2} S^-_{l+3}+{\rm H.c.}\;, \\
A^5 &=& \sum_l S^+_{l} S^-_{l+1} S^+_{l+2} S^-_{l+3}+{\rm H.c.}\;, \\
A^6 &=& \sum_l S^z_{l} S^+_{l+1} S^z_{l+2} S^-_{l+3}+{\rm H.c.}\;, \label{s32}
\end{eqnarray}
with $\mathcal{V}_{12} \simeq 0.4739$,  $\mathcal{V}_{22} = \mathcal{V}_{32} = \mathcal{V}_{42} = \mathcal{V}_{62} \simeq 0.3554$ and $\mathcal{V}_{52}\simeq-0.3554$. As before, the approximate LIOMs can be obtained via the rotation of $Q'^1_M$ and $Q'^2_M$ with respect to $Q^1_{M-2}\simeq H/\lVert H \rVert$. Figure \ref{fig3}(b) show results for $Q^2=Q^2_M$.

\subsection{Quasilocal integrals of motion}

\begin{figure}[thbp]
    \centering
    \includegraphics[width=\linewidth]{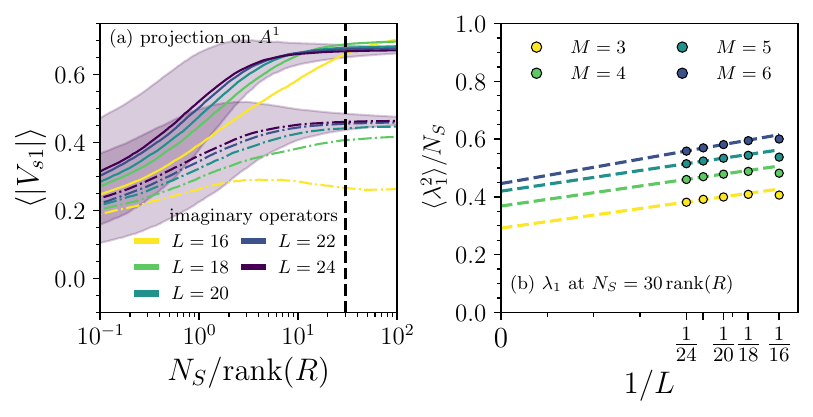}
    \caption{Results for the sector of imaginary operators that are odd under the spin-flip transformation. We consider the set of operators with the support $M=6$, for which \(\mathrm{rank}(R)= 66\). This symmetry sector includes the spin-current operator but does not contain any LIOMs. (a) The projections of $Q^1$ onto the operators defined in Eqs.~\eqref{s41}-\eqref{s42}. The shaded areas indicate the standard deviations for the largest ($L=24$) systems, using consistent color coding. (b) The largest singular value, $\langle \lambda_1^2 \rangle/N_S$, obtained for \(N_{S} = 30\, \mathrm{rank}(R) \). The extrapolation to $L\to\infty$ indicates that it increases with increasing support in the thermodynamic limit, and suggest that it corresponds to a QLIOM. }
    \label{fig4}
\end{figure}

We now verify whether  the existence of quasilocal integrals of motion (QLIOMs) can also be established from a vanishing fraction of all eigenstates. Unlike LIOMs, which are strictly local, QLIOMs have exponentially small but nonzero weights on operators acting over arbitrarily large distances~\cite{ilievski2015b, prosen13, mierzejewski15b,ilievski15}. Therefore, even when all eigenstates are taken into account (i.e., $N_s=Z$), QLIOMs cannot be reproduced exactly from the set of operators with an arbitrary finite support. This is reflected in the ratio $\lambda^2/Z$, which remains smaller than one for any finite $M$ and approaches unity only in the limit $M \to \infty$. Despite this, a QLIOM can still contribute to the Mazur bounds for local operators, affecting their dynamics and even preventing their thermalization. The requirement, which can also be regarded as a criterion for quasilocality, is that its projection onto some of the operators with finite support, $\mathcal{V}_{s\alpha}$, remains nonzero in the thermodynamic limit.

In order to detect QLIOMs, we have studied the sector of imaginary operators that are odd under the spin-flip transformation. This sector contains, for example, the spin current operator. At the same time, it does not contain LIOMs, so that the largest singular values obtained from Eq.~(\ref{svd}) may correspond to QLIOMs. Figure~\ref{fig4}(a) shows results for the largest projections of $Q^1$, which are found for the following local operators:
\begin{eqnarray}
A^1 &=& i \sum_l S^+_{l} S^-_{l+1} +{\rm H.c.}\;, \label{s41} \\
A^2 &=& i \sum_l S^+_{l} S^-_{l+2} +{\rm H.c.}\;. \label{s42}
\end{eqnarray}
The numerical calculations were performed for the set $\{A^1, \ldots, A^{D_O}\}$, which included operators with support on up to $M = 6$ sites.

Similarly to the cases discussed above, these projections, $V_{s1}$ with $s\le 2$, converge already for a small number of eigenstates, $N^*\simeq {\rm rank}({\cal R})\div10 \; {\rm rank}({\cal R})$. Since they do not decrease with $L$ and $\lim_{L\to \infty} V_{s1}={\cal V}_{s1}$, we expect that $Q^1$ is indeed a quasilocal operator. We also show how the corresponding singular value $\lambda_1$ depends on $L$ and $M$ in Fig.~\ref{fig4}(b). These results have been obtained for $N_s/{\rm rank(R)} =30$, which is marked by the dashed vertical line in the panel (a). While we are not able to obtain the accurate value of $\lim_{M\to\infty} \lim_{L\to\infty} \langle  \lambda^2_1 \rangle /N_s$, we observe that it is close to unity, as it should be for a conserved operator.

To conclude this section, we note that in the Bethe-ansatz integrable XXZ chain, LIOMs can be well estimated from just a tiny fraction of eigenstates. Moreover, the larger the system, the fewer eigenstates are needed.  This holds true for all considered symmetry sectors as well as for the detection of quasilocal integrals of motions.
We emphasize that for a fixed size, $L$, and support, $M$, taking into account more eigenstatestates does not yield any additional significant information regarding the existence or the structure of LIOMs or QLIOMs.

\section{Compression-based approach in fragmented system}
\label{sec:frag}

Hilbert-space fragmentation is an ergodicity-breaking phenomenon believed to be distinct from integrability~\cite{khemani17,sala2020,moudgalya21,moudgalya2022b,francica23}. However, many of the measures conventionally used to characterize closed quantum systems -- such as spectral statistics~\cite{oganesyan07,atas13,sierant19}, entanglement entropy~\cite{beugeling15,vidmar17,haque22,bianchi22,kliczkowski23}, matrix elements of local observables~\cite{jansen19,schonle21,richter20,wang22,dymarski22} and fidelity susceptibilities~\cite{pandey20,sels21,leblond21,kim24,swietek25,lisiecki25} -- exhibit very similar behavior in both cases. Consequently, it remains unclear whether, and if so, which properties fundamentally distinguish these two mechanisms of ergodicity breaking.

When written in a local basis (e.g., the computational basis), the Hamiltonian of a fragmented system shatters into exponentially many disconnected blocks.
For the dimension of the largest block, $Z_B$, one requires that 
$\lim_{L\to \infty} Z_B/Z=0$. Therefore, for a few randomly selected eigenstates, it is natural to expect that they all belong to different disconnected Krylov subspaces. To investigate whether this property affects the construction of LIOMs (whether it influences how information about the LIOMs is encoded in the eigenstates), we consider the folded XXZ model~\cite{pozsgay2021,zadnik21, zadnik21a}:
\begin{align}
   H =& \,\sum_{l} \left(\frac{1}{4} + S^z_{l-1}S^z_{l+2}\right)\left(S^+_{l}S^-_{l+1}+S^-_{l}S^+_{l+1}\right) \nonumber \\
   &+ g\sum_l \left( S^z_{l} S^z_{l+2} + S^z_{l} S^z_{l+3} \right).
   \label{folded}
\end{align}
The spin-flip term in Eq.~\eqref{folded} conserves the number of domain walls, 
\begin{equation}
A^1=\sum_l S^z_l S^z_{l+1}. \label{s5}
\end{equation}
Therefore, when $g=0$, the Hamiltonian from Eq.~\eqref{folded} emerges as an effective model of the XXZ chain in the limit of infinite $\Delta$, see Eq.~\eqref{xxz}. We have added an additional term $\sim g$ to be able to break the Bethe-ansatz integrability without destroying the fragmentation of the Hilbert space.
This term allows us to determine the origin of the anomalous behavior that emerges from the numerical studies, as demonstrated below.

The folded XXZ model has massively degenerate eigenstates. For this reason, one  needs to introduce a few  technical modifications to the method. Since we study  translationally-invariant operators, for which nonvanishing matrix elements exist only within the momentum sectors, we consider degeneracies only within such sectors. Since  degeneracies are present in all of them, we also take into account the $k=0$ and $k=\pi$ momentum sectors. The matrix $\cal R$ from Eq.~\eqref{rcal} as well as $R$ from Eq.~\eqref{svd} should always contain all matrix matrix elements from a degenerate subspace, i.e., these matrices should store (in columns) all $A^s_{nm}=\langle n |\hat A^s | m\rangle$ for $E_m=E_n$~\cite{lydzba2024}. In order to facilitate the numerical calculations for a massively degenerate eigenproblem, we use the exact denationalization instead of the shift-invert method. Finally, we note that a random drawing of a fixed number of eigenstates in the case of degeneracies is ambiguous. One would also seek to avoid a situation in which states belonging to degenerate subspaces are more likely to be included in the matrix $R$ than non-degenerate states. Therefore, instead of drawing $N_S$ eigenstates, we draw $N_D$ degenerate subspaces. In order to maintain consistency with the previous discussion, all numerical results will be presented in terms of
\begin{equation}
\tilde{N_s}=N_D \bar{Z}_D,   
\end{equation}
where $\bar{Z}_D$ is the average dimension of the degenerate subspaces calculated from all momentum sectors. We note that in the nondegenerate case, $\tilde{N_S}$ is identical to the previously considered $N_S$. Moreover, when all degenerate subspaces are taken into account, one obtains $\tilde{N}_S = Z$.

\begin{figure}[thbp]
    \centering
    \includegraphics[width=\linewidth]{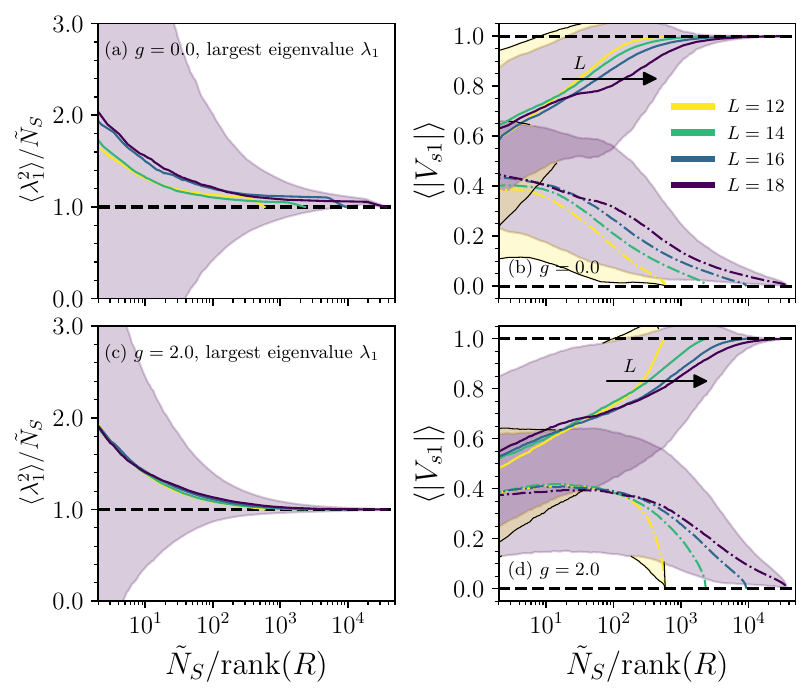}
    \caption{Results for the folded XXZ model from Eq.~\eqref{folded}. The set of $A^s$ is constructed only from $S^z_l$ and $I_l$ acting on up to \(M=4\) consecutive sites, for which \(\mathrm{rank}(R)=7\). All results are averaged over $10^4$ different choices of $N_S$ eigenstates, also in Fig.~\ref{fig6}.
    (a) The largest singular value, $\langle\lambda_1^2\rangle/\tilde{N_s}$, for $g=0$. (b) Top curves correspond to the largest projection of $Q^1$ onto the operator $A^1$ from Eq.~\eqref{s5}, while bottom curves correspond the next largest projection. The shaded areas indicate the standard deviations of the projections for the smallest ($L=12$) and largest ($L=18$) systems, using consistent color coding. The edge of the former is marked with a thin black curve for visibility. Panels (c) and (d) show the same results as panels (a) and (b), respectively, but for the perturbed model with \(g=2.0\). 
    }
    \label{fig5}
\end{figure}
\begin{figure}
    \centering
    \includegraphics[width=1.0\linewidth]{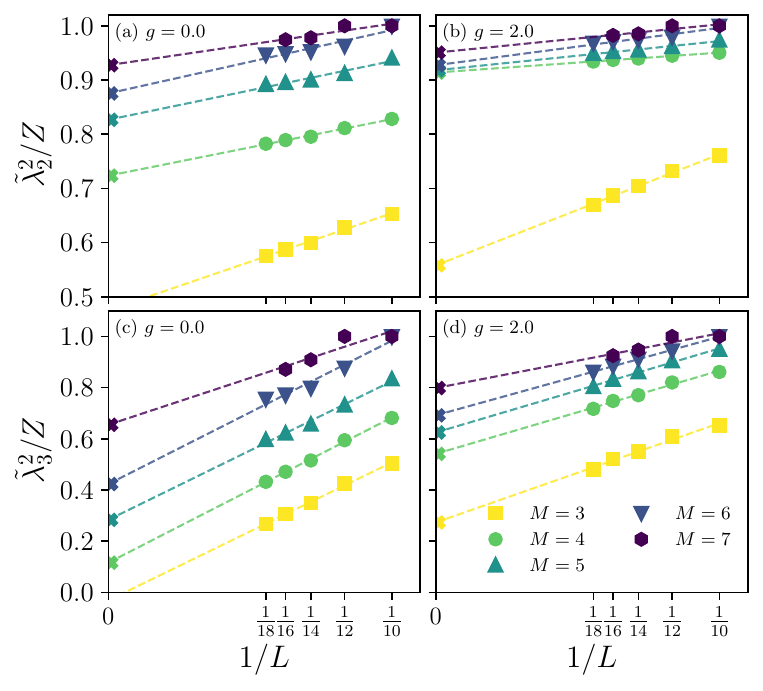}
    \caption{
    (a),(b) The second and (c),(d) third singular values in the folded XXZ model, which were obtained from the same set of operators as in Fig.~\ref{fig5} but for all eigenstates (i.e., $\tilde{N}_S=Z$). (a),(c) show the results for the unperturbed model with $g=0.0$, while (b),(d) show the results for a relatively strong perturbation with $g=2.0$.}
    \label{fig:folded_qlioms}
\end{figure}
\begin{figure}[thbp]
    \centering
    \includegraphics[width=\linewidth]{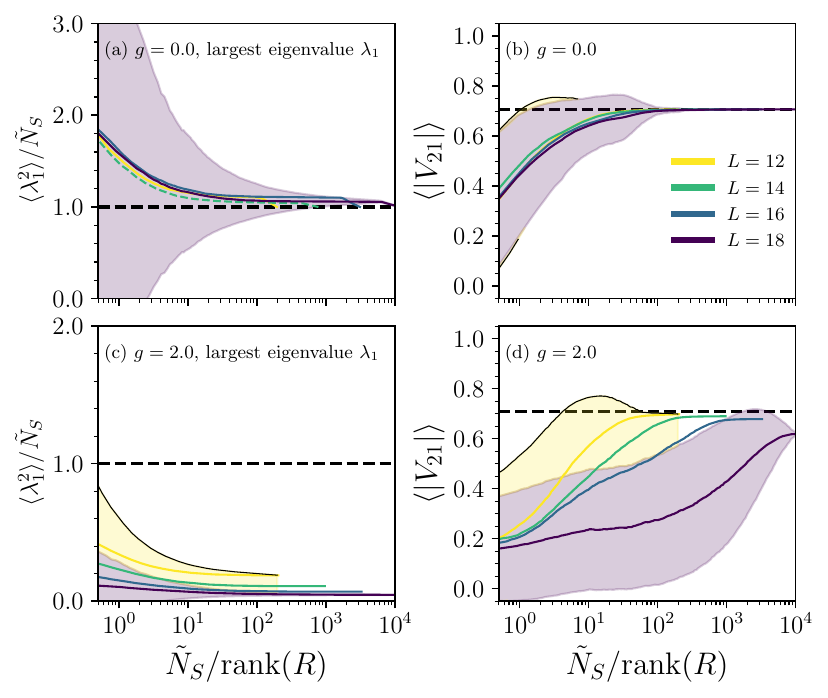}
    \caption{Results for the XXZ folded model, as in Fig.~\ref{fig5}, but for another set of operators. Here, we have used the same set as in Fig.~\ref{fig1}, for which \(M=4\) and \( \mathrm{rank}(R)=5\).
    (b),(d) show the largest projection of $Q^1$ on the operator $A^2$ from Eq~\eqref{s112}.
    }
    \label{fig6}
\end{figure}

We first apply the compression-based approach to a small set of operators, which are built only from $S^{z}_l$ and $I_l$ operators at each site, and supported on up to $M=4$ sites. For clarity, the trivially conserved total magnetization, $\sum_l S^z_l$, is excluded from this set.
It contains a LIOM, ${\cal Q}^1=A^1$ from Eq.~\eqref{s5}, which we aim to single out for $g=0$. Figure~\ref{fig5}(a) shows the largest singular value, $\langle\lambda_1^2\rangle/\tilde{N_s}$.
Simultaneously, Fig.~\ref{fig5}(b) demonstrates how the largest projection of ${Q^1}$ onto ${\cal Q}^1=A^1$ (continuous lines) and the second largest projection of ${Q^1}$ (dashed lines) scales with $\tilde{N_s}$. As expected, $\langle  \lambda^2_1 \rangle/\tilde{N_s}$ eventually approaches unity and the corresponding ${Q^1}$ eventually becomes ${\cal Q}^1$ for large $\tilde{N_s}$. Nevertheless, in order to accurately estimate this LIOM, a majority of eigenstates has to be used ($N^* \simeq Z$), and the larger the system, the more eigenstates are required. This is exactly the opposite of the behavior observed earlier for the LIOMs in the XXZ model (with a finite $\Delta$), cf.~Fig.~\ref{fig2}.

To unequivocally link ${\cal Q}^1$ with the Hilbert-space fragmentation, we have repeated the calculations while breaking the Bethe-ansatz integrability by setting $g=2$.
The corresponding results shown in Figs.~\ref{fig5}(c) and \ref{fig5}(d) are very similar to the results for the integrable case shown in Figs.~\ref{fig5}(a) and \ref{fig5}(b), respectively. This demonstrates that the conservation of ${\cal Q}^1$ is not related to the Bethe-ansatz integrability. 

Interestingly, our numerical results indicate that $\mathcal{Q}^1$ is not the only conserved quantity arising from Hilbert space fragmentation and persisting for nonzero $g$. In Fig.~\ref{fig:folded_qlioms}, we present the second and third largest singular values obtained for the same set of operators, which are built only from products of $S^{z}_l$ and $I_l$. In this particular case, we have used all eigenstates of the Hamiltonian from Eq.~\eqref{folded}, i.e., $\tilde{N_s}=Z$. 
Figure \ref{fig:folded_qlioms} shows how these singular values depend the system size $L$ and the support $M$.
The presence of QLIOMs is even more clear than in the previously discussed case, see Fig.~\ref{fig4}(b). In particular, $\lim_{M\to \infty}\lim_{L\rightarrow\infty} \tilde{\lambda}^2_{2,3}/Z$ approaches unity even more clearly. This is especially evident for the perturbed system with $g=2$ [see Fig.~\ref{fig:folded_qlioms}(b) and~\ref{fig:folded_qlioms}(d)], reinforcing the claim that these conservation laws do not originate from the Bethe-ansatz integrability. To determine the exact structure of these QLIOMs, additional calculations are required, which lie beyond the scope of the present work.

Finally, we consider another set of operators, which allows for the construction of LIOMs that originate from the Bethe-ansatz integrability. Specifically, we consider the same set of operators that has previously been used for the XXZ model with $\Delta=3/4$, for which results are shown in Fig.~\ref{fig1}. For the case of the folded XXZ model, the largest singular value, $\langle\lambda_1^2\rangle/\tilde{N_s}$, is shown in Fig.~\ref{fig6}(a). The corresponding $Q^1$ represents the energy current and its largest projection on the operator $A^1$ from Eq.~\eqref{s11} is plotted in Fig.~\ref{fig6}(b). In order to connect the obtained integral of motion to the Bethe-ansatz integrability, we have repeated the same calculations but for $g=2$, see Figs.~\ref{fig6}(c) and \ref{fig6}(d). It is clear that when $g$ is different from zero, the same set of operators does not contain any LIOM. 
The comparison of Figs.~\ref{fig6}(a) and \ref{fig6}(b) with Figs.~\ref{fig1}(d) and \ref{fig1}(a), respectively, shows that Hilbert-space fragmentation does not prevent the construction of LIOMs, provided that these LIOMs do not originate from it. Even in its presence, such LIOMs can be reliably estimated from $N^* \simeq {\rm rank}(R) \div 10{\rm rank}(R)$ eigenstates.

Since the energy current can be reconstructed from a small number of eigenstates, $N^*\ll Z$, the inability to do the same for LIOMs originating from the Hilbert space fragmentation does not appear to be merely a consequence of massive degeneracies. One might expect that the information about LIOMs is not contained in individual eigenstates within a degenerate subspace, but is instead somehow distributed among them. In such a case, including a single degenerate subspace is more akin to including a single non-degenerate eigenstate. Note that for $L=16$, when the Hilbert space dimension is $65536$, there are $8314$ (momentum-resolved) degenerate subspaces with an average dimension of $8$, while the maximal dimension of $665$. Yet, the observed inability appears to reflect a more subtle property of fragmented systems.

\section{Concluding remarks}
\label{sec:con}

In this work, we have demonstrated that integrability leaves a significant imprint on individual eigenstates, enabling the reconstruction of local integrals of motion (LIOMs) from an incomplete spectrum. This greatly broadens our understanding of the information encoded in individual eigenstates. Specifically, their small number not only allows the identification of ergodic systems and the estimation of their thermal properties, but also makes it possible to establish local conservation laws in integrable systems and, consequently, to estimate the properties of their stationary states.

We construct LIOMs by compressing information contained in the diagonal (or, more generally, equal-energy) matrix elements of local observables. This compression problem is well defined even for an incomplete set of eigenstates. Studying the XXZ chain, which is integrable, we find that LIOMs can be constructed from a set of $D_O$
 local operators using a number of eigenstates only slightly larger than $D_O$. This number does not grow with the system size, and the variations across different choices of eigenstates remain small. These properties hold for LIOMs in different symmetry sectors as well as for quasilocal integrals of motion.
 
In contrast, the described properties are violated in the folded XXZ model, which exhibits the Hilbert-space fragmentation. The model exhibits conserved quantities that are unrelated to Bethe-ansatz integrability, since they remain robust under perturbation that breaks integrability but preserves fragmentation. Interestingly, constructing these conserved quantities requires access to almost the entire set of eigenstates. We argue that this cannot be attributed solely to the massive degeneracies present in the model. This finding represents one of the few fundamental differences known between integrability and fragmentation.

\begin{acknowledgments}
We acknowledge fruitful discussions with Lev Vidmar.
This work was supported by the National Science Centre (NCN), Poland via project 2023/49/N/ST3/01033.
\end{acknowledgments}

\bibliography{manupert}

\end{document}